\documentclass[11pt]{article}

\usepackage{latexsym,amsfonts,amsmath,amssymb,wasysym,enumerate,epsfig}
\usepackage{theorem}
\usepackage{mathrsfs}
\usepackage{tikz}
\usepackage{ulem}

\usepackage[all]{xy}

\usepackage{color}

\numberwithin{equation}{section}

\newcommand{\mb}[1]{\quad\mbox{#1}\quad}
\newcommand{\be}{\begin{equation}}
\newcommand{\ee}{\end{equation}}
\newcommand{\beu}{\begin{equation*}}
\newcommand{\eeu}{\end{equation*}}
\newcommand{\bea}{\begin{eqnarray}}
\newcommand{\eea}{\end{eqnarray}}
\newcommand{\beaa}{\begin{eqnarray*}}
\newcommand{\eeaa}{\end{eqnarray*}}
\newcommand{\bmx}{\begin{pmatrix}}
\newcommand{\emx}{\end{pmatrix}}

\newcommand{\p}{{\mathfrak p}}

\newcommand{\wh}[1]{{\widehat{#1}}}

\def\with{{\quad\mbox{with}\quad}}

\def\and{\quad\mbox{and}\quad}

\newcommand{\llangle}{\langle\!\langle}
\newcommand{\rrangle}{\rangle\!\rangle}
\newcommand{\steady}{|{\cal S}\rangle} 
\advance\textheight by \topskip
\begin{document}
\setcounter{page}{0}
\pagestyle{empty}
%
%
\begin{center}

 {\LARGE  {\sffamily  Open two-species exclusion processes\\[1.2ex]
 with integrable boundaries} }\\[1cm]

\vspace{10mm}
  
{\Large 
 N. Crampe$^{a}$\footnote{nicolas.crampe@univ-montp2.fr}, K. Mallick$^b$\footnote{kirone.mallick@cea.fr},
 E. Ragoucy$^{c}$\footnote{eric.ragoucy@lapth.cnrs.fr}
 and M. Vanicat$^{c}$\footnote{matthieu.vanicat@lapth.cnrs.fr}}\\[.41cm] 
{\large $^a$ Laboratoire Charles Coulomb (L2C), UMR 5221 CNRS-Univ. Montpellier 2,\\[.242cm]
Montpellier, F-France.}
\\[.42cm]
{\large $^{b}$   Institut de Physique  Th\'eorique\\[.242cm] 
 CEA Saclay, F-91191 Gif-sur-Yvette, France. }
\\[.42cm]
{\large $^{c}$ Laboratoire de Physique Th{\'e}orique LAPTh
 CNRS and Universit{\'e} de Savoie.\\[.242cm]
   9 chemin de Bellevue, BP 110, F-74941  Annecy-le-Vieux Cedex, 
France. }
\end{center}
\vfill

\begin{abstract}
We give a complete classification of integrable Markovian boundary conditions
for the  asymmetric simple exclusion process with two species (or
classes) of particles.  Some of these  boundary conditions  lead to
non-vanishing particle  currents for each species.  We explain how the
stationary state of all these models  can  be expressed in a matrix
product form, starting from two key components,  the
Zamolodchikov-Faddeev and  Ghoshal-Zamolodchikov relations.
This  statement is illustrated  by studying in detail a specific
example,  for which the matrix Ansatz (involving 9 generators)  is
explicitly constructed and physical observables (such as currents,
densities) calculated. 
\end{abstract}

\vfill\vfill
\rightline{LAPTh-237/14}
\rightline{December 2014}

\newpage
\pagestyle{plain}

\section{Introduction}

 Particles  that diffuse  on a discrete  lattice, with anisotropic
 hopping rates and  hard-core exclusion,  define one of the simplest
 examples of a driven diffusive system \cite{KLS,PaulK,Zia}, known as
 the asymmetric simple exclusion process (ASEP).  Thanks to  its
 simplicity, the ASEP plays  a seminal role  in  classical
 low-dimensional transport with excluded volume constraints  and
 appears in numerous  phenomenological models  in hard and soft
 condensed matter (see e.g. \cite{CKZ,DerridaRep,Schutz} and
 references therein).  The exclusion process is also a paradigm in the
 field of  non-equilibrium statistical mechanics
 \cite{Bertini,DerrReview,Liggett2,Spohn91}:  it displays  an
 unexpectedly complex  behaviour  \cite{Varadhan},  that has
 stimulated  many  elaborate analytical studies.  A distinctive
 feature of the ASEP  (as compared to many other interacting particles
 models  such as the zero range process) is  that it  is  {\it
 integrable.}  This remarkable property  explains  its tremendous
 success  in mathematical physics, combinatorics, representation
 theory and  probability theory
 \cite{Alcaraz1,Corwin,Duchi,Isaev,Krug2,Sasamoto1}. Besides,  many intricate
 questions about non-equilibrium behaviour,    when formulated for the
 ASEP, allow for precise and  quantitative answers, often involving
 elegant mathematical structures. Hence,  the ASEP  provides  us with
 benchmarks for a  general theory of non-equilibrium systems
 \cite{GianniRevue,DerrReview,DCairns}.

 In the   exclusion process, identical particles  undergo a stochastic
 evolution on a lattice;  empty sites are called {\it holes}. A
 natural generalization  of the original dynamics  involves multiple
 species (or classes) of particles; second class  particles were
 initially used  as a tool to locate  the shock  that may  appear at
 the microscopic level in the  density profile  of the ASEP
 \cite{DJLS,DLS,Ferrari,Ferrarietal}.  More generally,  introducing
 multiple species of particles is  a way to  couple  independent  ASEP
 models, a standard probabilistic technique  that allows to gain
 insight on individual processes \cite{Liggett2, Thorisson}. Moreover,
 the  ASEP with multiple species  has some  physical applications,
 e.g. in relation to traffic flow \cite{MartinRev,Evans,Karimipour}.

 The steady state distribution of the two-species exclusion process on
 a ring was obtained  in \cite{DJLS} by using the matrix product
 representation introduced by B.  Derrida et al.  in
 \cite{DEHP}. This exact  solution of the two-species exclusion
 process had multiple outcomes: shock profiles were   calculated
 analytically \cite{DJLS}; the invariant measure was proved to be
 non-Gibbsean \cite{Speer}; quadratic algebras involved in the
 `matrix Ansatz' were  studied in
 \cite{Aneva,Essler,Isaev}; combinatorial  interpretations of the
 weights generated by the  matrices  were found
 \cite{Angel,Duchi,FerrariFontes} etc.  These results  were  extended
 to models  with an impurity  \cite{km} and to   systems defined on
 the infinite  line  \cite{DLS}  (many references can be found in the
 standard review \cite{MartinRev}).  The solution of the exclusion
 process with three or more species  on a periodic or on the infinite
 lattice turned out to be more involved; it relies on an  astute
 combinatorial construction \cite{FerrariMartin1,FerrariMartin2}  (see
 also \cite{Ayyer,Linusson}) that can be encoded by a  tensor matrix
 Ansatz \cite{EFM,MMR,PEM}. These multi-species exclusion processes
 remain  integrable and  systems defined on a periodic lattice can, in
 principle, be solved by a  nested Bethe Ansatz
 \cite{Alcaraz2,Arita2,Arita3,Cantini,DEvansImp}.

 Models  defined on a finite lattice  with open boundaries   connected
 to  reservoirs, that allow for the exchange of particles, represent a
 major challenge. From the physicist's point of view, these models are
 more realistic and therefore the most interesting.  Indeed, the
 fundamental picture often used to illustrate a non-equilibrium system
 is a pipe connecting two (or more) reservoirs at different
 temperatures, or chemical potentials. The discovery of
 boundary-induced phase transitions in the open exclusion process
 \cite{Krug}, followed by the  matrix Ansatz  solution  \cite{DEHP}
 were the  seminal contributions that did trigger the whole
 field. However,  for the case of  the  multi-species ASEP  with open
 boundaries  many question  remain to be clarified \cite{Arita0,Arita1,ALS1,ALS2,DEvans,Uchi} (see section \ref{previous_works} for 
  the state of the art).
 
In the present work, we wish to investigate the asymmetric  exclusion
process with different species of particles.  This model,  which
will be called in brief  the   `2-ASEP', is   defined on a finite
lattice of $L$ sites where  boundary sites  are connected to reservoirs.
The precise dynamical rules of the 2-ASEP will be given in Section~\ref{sec:summary}.
At the first  and  the last
sites, particles  can either enter or exit: thus, 
at  each boundary, there are 
6 possible choices for the corresponding  rates.
 The aim of this
 work is to give a complete classification of the boundary
 rates that lead to integrable models. In other words, we find the
 conditions on the boundary rates that preserve the  integrability of
 the 2-ASEP. This is done by solving simultaneously the Yang-Baxter
 equation in the bulk and the Sklyanin reflection equations at  the
 boundaries \cite{CRV,Faddeev,sklyanin}.   The
 models thus obtained, are the only ones solvable by Bethe Ansatz and we are convinced  that they are also 
 analytically tractable by Matrix Ansatz \cite{CRV}. 

The outline of this work is as follows. In section \ref{sec:summary},
we present all the dynamical rules for the  2-ASEP with open
boundaries that preserve integrability; we also discuss the relation
between these integrable  models and  previous studies.   In section
\ref{int_sys}, we apply  the theory of integrable systems to  the open
2-ASEP, starting from  the  (bulk) $R$-matrix and two boundary $K$
matrices.  The operators $R$ and $K$ have to satisfy  non-linear
algebraic relations (the  Sklyanin reflection  equations) for which all
solutions are found and classified. Each solution  allows us to define
a one-parameter family of commuting transfer matrices, which yields,
finally, to an integrable  Markov operator of the 2-ASEP with open
boundaries.  In the subsection \ref{sec:MA}, we explain, using the
results of  \cite{CRV},  how a matrix Ansatz  can be derived from this
formalism. Section \ref{example} is devoted to a specific example,
perhaps the simplest 2-TASEP model with open boundaries 
 in which all types of particles can enter and leave the system.
 We construct a Matrix Ansatz for this model: the underlying algebra 
  contains nine generators and  differs from 
  the quadratic algebra  that solves  the single species ASEP
 with open boundaries   (hereafter, we shall refer to this quadratic algebra,
  introduced  in \cite{DEHP}, as the DEHP algebra).
The last section is
devoted to some concluding remarks.

\section{Integrable dynamics for  the open  2-ASEP}
\label{sec:summary}

\subsection{The ASEP case}
  We start  by  recalling  the dynamical rules of the 2-ASEP.  Each
  site on the lattice is occupied by a single particle, and there are
  three different types (or classes) of particles, denoted by 3, 2 and
  1.  The occupation variable at a site $i$ of  the system, with   $ 1
  \le i \le L$, is denoted by $\tau_i$: we have  $\tau_i=1,2$ or 3.
  The bulk dynamics is defined as follows: a  bond $(i,i+1)$,  with $
  1 \le i \le L-1$, between two neighboring lattice sites, is updated
  between time $t$ and $t +dt$ by swapping the particles  at $i$ and
  $i+1$; if  $\tau_i > \tau_{i+1}$,  the exchange occurs with rate 1,
  whereas if  $\tau_i < \tau_{i+1}$ the exchange occurs with rate $q
  <1$.

 These bulk rules are summarized in the following table: 
\begin{eqnarray}
\begin{array}{|c|c|} 
\hline
\hspace{0.5cm} 21\, \xrightarrow{\ 1\ } \,12\hspace{0.5cm}&\hspace{0.5cm} 12\, \xrightarrow{\ q\ }\, 21\hspace{0.5cm}\\
\hspace{0.5cm} 31\, \xrightarrow{\ 1\ }\, 13\hspace{0.5cm}&\hspace{0.5cm} 13\, \xrightarrow{\ q\ }\, 31\hspace{0.5cm}\\
\hspace{0.5cm} 32\, \xrightarrow{\ 1\ } \,23\hspace{0.5cm}&\hspace{0.5cm} 23\, \xrightarrow{\ q\ }\, 32\hspace{0.5cm} \\
\hline
 \end{array}
\end{eqnarray}

   These rules show that particles of type 3  have the highest
   priority, followed by 2's and lastly by 1's (we emphasize that our
   notation differs from that  of some previous works
   \cite{EFM}. Type 3 particles  having highest priority were called
   first class particles, type 2 second class and 1's play the role of
   holes). These bulk rules define an integrable dynamical system,
   that can be encoded by a  $R$-matrix, that  satisfies  the
   Yang-Baxter equation \cite{Alcaraz2,Arita2,Faddeev}.  The R-matrix
   will be given explicitly in the next section. 

 Particles are allowed to enter or to exit from both boundaries and
 the corresponding  entrance/exit rates  may depend on the type of the
 particle that was previously located at the boundary. More precisely,
 both on the left and on the right boundary, we can have  a transition
 of the type $I \rightarrow J$, where $I$ and $J$ represent particles
 of different classes, i.e.  $ 1 \le I,J \le 3$ and $I \neq J$. This
 leads to 12 independent rates (6 on each side). However, for 
 arbitrary choices of these rates, the models will not be integrable.

  Hence, in order to  preserve  integrability, these 12
  rates cannot be chosen in an arbitrary manner.  They have to
  satisfy integrable boundary conditions that are associated to the
  bulk $R$-matrix. These  boundary conditions  are expressed as a set
  of reflection equations \cite{sklyanin}, 
  which will be solved in Section~\ref{int_sys} and their solutions, the
  K-matrices, which will be  fully determined. 

     From this  theory of integrable systems  with boundaries, we  shall derive
  four sets of integrable rules at the  left
boundary:  out of the six possible rates, only two are independent,
the other ones either vanish or are simple functions of the two
independent rates.  Calling by  $\alpha$ and $\gamma$ two arbitrary
positive real numbers, we find that the four set of left-rules are
given by
\begin{eqnarray}\label{rate_left} 
\begin{array}{|c|c|c|c|} 
\hline 
L_1&L_2&L_3&L_4\\ 
\hline
\rule{0ex}{3.7ex}\hspace{0.5cm} 1\, \xrightarrow{\ f(\alpha,\gamma)\ }\, 2\hspace{0.5cm}&\hspace{0.5cm} 1 \,\xrightarrow{\quad \ \gamma\ \quad}\, 2\hspace{0.5cm}&\hspace{0.5cm} 1 \,\xrightarrow{\ \alpha\ }\, 3\hspace{0.5cm}&\hspace{0.5cm} 1 \,\xrightarrow{\ \alpha\ }\, 3 \hspace{0.5cm}\\
\hspace{0.5cm} 2 \,\xrightarrow{\quad \ \alpha \quad }\, 1\hspace{0.5cm}&\hspace{0.5cm} 1 \,\xrightarrow{\quad \ \alpha \quad }\, 3\hspace{0.5cm}&\hspace{0.5cm} 2 \,\xrightarrow{\ \gamma\ }\, 1\hspace{0.5cm}&\hspace{0.5cm} 3 \,\xrightarrow{\ \gamma\ }\, 1\hspace{0.5cm}\\
\hspace{0.5cm} 3 \,\xrightarrow{\quad \ \alpha \quad } \,1\hspace{0.5cm}&\hspace{0.5cm} 2 \,\xrightarrow{\quad \ \alpha \quad }\, 3\hspace{0.5cm}&\hspace{0.5cm} 2 \,\xrightarrow{\ \alpha\ }\, 3\hspace{0.5cm}& \\
\hspace{0.5cm} 3 \,\xrightarrow{\quad \ \gamma \quad } \,2\hspace{0.5cm}&\hspace{0.5cm} 3 \,\xrightarrow{\ g(\alpha,\gamma)\ }\, 2\hspace{0.5cm}&\hspace{0.5cm} 3 \,\xrightarrow{\ \gamma\ }\, 1\hspace{0.5cm}&  \\
\hline
 \end{array}
\end{eqnarray}
where we have defined 
\begin{equation}
 f(x,y)=\frac{y(x+y+1-q)}{x+y} \ \text{ and } \ g(x,y)=\frac{y(x+y+q-1)}{x+y}.
\label{def:fetg}
\end{equation}
{\it Remark:}  Note that in the $L_2$ case, we must have $\alpha + \gamma > 1 -q$ to ensure  $g(\alpha,\gamma) >0$

\hfill\break

Similarly, we obtain   four integrable rules at  the right boundary that can be encoded in the following table

\begin{eqnarray}\label{rate_right}
\begin{array}{|c|c|c|c|} 
\hline
R_1&R_2&R_3&R_4\\ 
\hline
\rule{0ex}{3.7ex}\hspace{0.5cm} 1\, \xrightarrow{\quad \ \delta \quad }\, 2\hspace{0.5cm}&\hspace{0.5cm} 1 \,\xrightarrow{\ g(\beta,\delta)\ }\, 2\hspace{0.5cm}&\hspace{0.5cm} 1 \,\xrightarrow{\ \delta\ }\, 3\hspace{0.5cm}&\hspace{0.5cm} 1 \,\xrightarrow{\ \delta\ }\, 3\hspace{0.5cm} \\  
\hspace{0.5cm} 1 \,\xrightarrow{\quad \ \beta \quad }\, 3\hspace{0.5cm}&\hspace{0.5cm} 2 \,\xrightarrow{\quad \ \beta \quad }\, 1\hspace{0.5cm}&\hspace{0.5cm} 2 \,\xrightarrow{\ \beta\ }\, 1\hspace{0.5cm}&\hspace{0.5cm} 3 \,\xrightarrow{\ \beta\ }\, 1\hspace{0.5cm}\\ 
\hspace{0.5cm} 2 \,\xrightarrow{\quad \ \beta \quad } \,3\hspace{0.5cm}&\hspace{0.5cm} 3 \,\xrightarrow{\quad \ \beta \quad }\, 1\hspace{0.5cm}&\hspace{0.5cm} 2 \,\xrightarrow{\ \delta\ }\, 3\hspace{0.5cm}& \\ 
\hspace{0.5cm} 3 \,\xrightarrow{\ f(\beta,\delta)\ } \,2\hspace{0.5cm}&\hspace{0.5cm} 3 \,\xrightarrow{\quad \ \delta \quad }\, 2\hspace{0.5cm}&\hspace{0.5cm} 3 \,\xrightarrow{\ \beta\ }\, 1\hspace{0.5cm}&   \\
\hline
 \end{array}
\end{eqnarray}
where $\beta$ and $\delta$ are two positive real numbers and $f(x,y)$, $g(x,y)$
 were  defined in (\ref{def:fetg}).

\subsection{The TASEP case}

 We  emphasize  here  the  TASEP limit ($q\to0$). For the TASEP,
 the bulk dynamics   is given by:
\begin{eqnarray}
  21\, \xrightarrow{\ 1\ } \,12\mb{ ; }
 31\, \xrightarrow{\ 1\ }\, 13\mb{ ; }
 32\, \xrightarrow{\ 1\ } \,23.
 \end{eqnarray}

 The integrable boundary transition rates compatible with the 2-TASEP
dynamics are obtained  by taking the limit $q\to0$ in
\eqref{rate_left} and \eqref{rate_right}. However, this limit allows
 for some `unphysical'  transitions, where particles can be
 injected or extracted in the direction
 opposite to  the natural orientation defined by  the bulk. In order to avoid
 these situations, we have to  fix the values of the parameters
$\gamma$ and $\delta$. We take $\gamma=1-\alpha$ and  $\delta=1-\beta$
 for $L_2$ and $R_2$ (so that $g(\alpha,\gamma)=0$ and $g(\beta,\delta)=0$), and  $\gamma=\delta=0$  for boundaries of type $L_3,L_4,$
 or $R_3,R_4$. Besides, in the cases $L_1$ and $R_1$, we need 
first to rescale $\alpha=\alpha' q$, $\gamma=\gamma' q$, $\beta=\beta' q$ and $\delta=\delta' q$ and then 
 take the limit $q\to0$.

Altogether, we end up with  four sets of integrable transition rates on the left boundary:
\begin{eqnarray}\label{rate_left_t}
\begin{array}{|c|c|c|c|}  \hline
L_1&L_2&L_3&L_4\\
\hline
\rule{0ex}{3ex}\hspace{0.5cm} 1 \,\xrightarrow{\ \mu\ }\, 2\hspace{0.5cm}&\hspace{0.5cm} 1 \,\xrightarrow{\ 1-\alpha\ }\, 2\hspace{0.5cm}&\hspace{0.5cm} 1 \,\xrightarrow{\ \alpha\ }\, 3\hspace{0.5cm}&\hspace{0.5cm} 1 \,\xrightarrow{\ \alpha\ }\, 3 \hspace{0.5cm}\\
&\hspace{0.5cm}1 \,\xrightarrow{\ \ \alpha \ \ }\, 3\hspace{0.5cm}&\hspace{0.5cm} 2 \,\xrightarrow{\ \alpha\ }\, 3\hspace{0.5cm}& \\
\hspace{0.5cm}\mu=\frac{\gamma'}{\alpha'+\gamma'}\hspace{0.5cm}&\hspace{0.5cm} 2 \,\xrightarrow{\ \ \alpha \ \ }\, 3\hspace{0.5cm}& &  \\
\hline
 \end{array}
\end{eqnarray}
There are also four sets of integrable transition rates on the right boundary:
\begin{eqnarray} \label{rate_right_t}
\begin{array}{|c|c|c|c|} 
\hline
R_1&R_2&R_3&R_4\\
\hline
\rule{0ex}{3.4ex}\hspace{0.5cm}3 \,\xrightarrow{\ \nu\ }\, 2\hspace{0.5cm}&\hspace{0.5cm}  2 \,\xrightarrow{\ \ \beta \ \ }\, 1\hspace{0.5cm}&\hspace{0.5cm} 2 \,\xrightarrow{\ \beta\ }\, 1\hspace{0.5cm}&\hspace{0.5cm} 3 \,\xrightarrow{\ \beta\ }\, 1\hspace{0.5cm}\\
&\hspace{0.5cm} 3 \,\xrightarrow{\ \ \beta \ \ }\, 1\hspace{0.5cm}&\hspace{0.5cm} 3 \,\xrightarrow{\ \beta\ }\, 1\hspace{0.5cm}& \\
\hspace{0.5cm}\nu=\frac{\delta'}{\beta'+\delta'}\hspace{0.5cm}&\hspace{0.5cm} 3 \,\xrightarrow{\ 1-\beta \ }\, 2\hspace{0.5cm}& &  \\
\hline
 \end{array}
\end{eqnarray}
The transition rates can be chosen among this four classes
independently for each boundary so that we have at the end
$4\times4=16$ different integrable models. The local boundary
operators and the corresponding  $K$-matrices will be  given in  the
next section.

\hfill\break
\subsection{Comparison with previous works\label{previous_works}}

   We now relate the integrable boundary conditions that we have found
   to  previously known results about  the 2-TASEP with open boundaries.
   In \cite{Arita0,Arita1}, C. Arita found a matrix Ansatz for the
   non-equilibrium stationary state (NESS) of  a 2-TASEP, using a
   variant of the quadratic algebra of \cite{DEHP}.  His solution was
   then generalized to the  partially asymmetric case \cite{Uchi}.
   The structure of this  NESS, the density profiles, the existence of
   { fat shocks} and some  mathematical properties of the invariant
   measures were thoroughly investigated by A. Ayyer at al.  \cite{ALS1}. These authors
    emphasized that all the models hitherto considered had the
   very special property that the boundaries were {\it semi-permeable}:
   second class particles (2) can neither enter nor leave the system. 
   These semi-permeable boundaries correspond to 
    our $L_4$ and $R_4$.
   This is a drastically restrictive condition: it implies, in particular,
   that the current $J_2$ of second class particles vanishes. In fact, it was shown in
    \cite{ALS1} that the distribution of these second class particles is given by
    an equilibrium ensemble in fixed volume.  We emphasize that the 
    integrable boundary conditions that we have found, are not restricted solely
    to semi-permeable boundaries. Some of the models we defined
    are genuinely out of equilibrium, in all respects.

  In a second work \cite{ALS2}, A. Ayyer et al. considered more general classes
 of open two-species exclusion processes. They showed that a matrix solution
 based on a quadratic algebra could also  be used for some permeable boundaries,
 if the rates obey some particular relations. The set of solvable 2-TASEP models
 was thus  extended. However, these authors also proved that although
 the boundaries were permeable, the current $J_2$ still identically vanishes in these
 models (in the limit of large system sizes).
 It seemed that  all  models, solvable  by a quadratic
 DEHP algebra with 3 generators, similar to  the ones introduced in \cite{DEHP},
  had  $J_2 =0$: this corresponds to the choices $L_4/R_2$, $L_2/R_4$, $L_4/R_1$, $L_1/R_4$, $L_2/R_2$ and $L_1/R_1$.
 
Note that for the choices $L_4/R_3$, $L_3/R_4$ and $L_3/R_3$ (resp. $L_1/R_2$, resp. $L_2/R_1$), the density of 2 (resp.  3, resp. 1)
 identically vanishes for all system sizes. The stationary state can be computed using exactly the same algebra as in \cite{DEHP}.
  
 Therefore, we are left with 4  possible integrable boundaries $L_2/R_3$, $L_3/R_2$, $L_3/R_1$ and $L_1/R_3$ with non-vanishing currents for each species of particles. Building a matrix Ansatz for these models requires a more complicated algebra, using  the construction introduced in \cite{CRV}. 
 In Section~\ref{example},  we shall analyze an  example,
 that  cannot be solved by a DEHP algebra:  it requires
 at least 9 generators (or a more complex tensor-product construction 
 based on a  3 generators algebra \cite{CMRV}).
 
   Finally, in  \cite{ALS2}  other types of models,  called {\it
     colorable}, were  studied;  these models are not necessarily
   solvable (e.g., a matrix representation for the steady state does
   not exist  in general)  but currents and density profiles can be
   easily extracted from the known results for  the single species
   TASEP.  The idea is to identify two different classes of particles
   (by painting  them with the same color)   and reduce the 2-TASEP to
   the original TASEP.  There are two possible colorings:  either we
   group the 1's  and the 2's  together (and treat them as holes) and
   leave the particles 3 alone; or we group the  2's  and  the 3's
   (colored particles)  and leave the 1's alone (holes). Both
   colorings are compatible with bulk dynamics; the question is
   whether the boundary conditions also permit these identifications.
   The boundary conditions $L_2, L_3$ and  $R_2,R_3$ are
   colorable. Therefore, a  model with  these boundaries conditions
   is integrable (by construction) and the rates $\alpha,\beta$ that
   appear in \eqref{rate_left_t} and  \eqref{rate_right_t} can be
   chosen so that none of the currents vanishes: there exists  2-ASEP
   models with open boundaries  which are exactly solvable and in
   which all particles  genuinely behave out of equilibrium. 

\section{Commuting transfer matrices  for the 2-ASEP}
 \label{int_sys}

   In this section, we recall the integrability properties of the 2-ASEP and explain how the boundary
 dynamical rules described in section~\ref{sec:summary}  are derived. 
\subsection{Integrability of the open 2-ASEP}
  A Markov process  can be defined by  a master equation
$$ \frac{d |P_t\rangle}{ dt}  =M|P_t\rangle$$ 
where $|P_t\rangle$ is the vector containing the probabilities of the different configurations of the system
at time $t$ and $M$ is the Markov  matrix. For the exclusion process
 only local jumps are permitted and $M$  can be written as:
\begin{equation} \label{eq:Markov}
 M=B_1 + \sum_{\ell=1}^{L-1} w_{\ell,\ell+1} + \bar B_L.
\end{equation}
The operator $M$ acts in the configuration space $\left(\mathbb{C}^3\right)^{\otimes L}$, 
where each component of the tensor space stands for a site on the lattice. The subscripts in 
\eqref{eq:Markov} indicate in which component of the tensor space the operators are acting on non trivially.
Each  $w_{\ell,\ell+1}$ is a $9$ by $9$ matrix that  encodes  the dynamics in the bulk (i.e. the update rules
 on the bond $({\ell,\ell+1})$). 
 The matrix $B$ (respectively $\bar B$) is $3$ by $3$ and  gives 
the dynamics on the left (respectively right) boundary. 

 This stochastic  model is said to be integrable when the Markov  matrix $M$ can be embedded in a one-parameter 
 family of commuting operators $t(z)$,
 called  transfer matrices. A systematic method of constructing  
 integrable models with open boundaries was given by Sklyanin
 \cite{sklyanin}. It rests on three algebraic objects: the  (bulk) $R$-matrix and two boundary matrices
  $K$ and $\widetilde K$. 

 The $R$-matrix obeys the Yang-Baxter equation \cite{Faddeev}:
\begin{equation}
\label{YBE}
 R_{12}\left(\frac{x_1}{x_2}\right)\,  R_{13}\left(\frac{x_1}{x_3}\right)\, R_{23}\left(\frac{x_2}{x_3}\right)=
 R_{23}\left(\frac{x_2}{x_3}\right)\,R_{13}\left(\frac{x_1}{x_3}\right)\,R_{12}\left(\frac{x_1}{ x_2}\right)\,.
\end{equation}
The matrix $K$  has to satisfy the so-called reflection equation \cite{sklyanin}:
 \begin{equation}
\label{eq:re}
 R_{12}\left(\frac{x_1}{x_2}\right)\, K_1(x_1)\, R_{21}\left(x_1 x_2\right)\, K_2(x_2)=
 K_2(x_2)\,R_{12}\left(x_1 x_2\right)\,K_1(x_1)\,R_{21}\left(\frac{x_1}{x_2}\right)\;.
\end{equation}
 Finally, the matrix $\widetilde K$ obeys a dual-reflection equation: 
\begin{equation}
\label{eq:dualre}
  \widetilde{K_2}(x_2)\left(\,R_{21}^{t_1}(x_1 x_2)^{-1} \right)^{t_1} \,  \widetilde{K_1}(x_1)
 \, R_{21}\left(\frac{x_2}{x_1}\right) =  R_{12}\left(\frac{x_2}{x_1}\right)\, \widetilde{K_1}(x_1) \,
\left(\,R_{12}^{t_2}(x_1 x_2)^{-1} \right)^{t_2} \,  \widetilde{K_2}(x_2) 
\end{equation}
where $t_1$ and $t_2$ represent transpositions with respect to the first and the second space,
 respectively.

\hfill\break
{\it Remark:}  $\widetilde{K}(x)$ can be constructed  \cite{CRV}  from an auxiliary operator 
 $\bar K(x)$, using the relation 
\begin{equation} \label{eq:K_tilde}
 \widetilde{K_1}(x) =
tr_0\left(\bar K_0\left(\frac{1}{x}\right)\,  ((R_{01}(x^2)^{t_1})^{-1})^{t_1} \, P_{01}\right).
\end{equation}
This  auxiliary  $\bar K(x)$ obeys  the  reflection equation \eqref{eq:re} associated
 with the matrix $\bar R=R^{-1}$:
 \begin{equation} \label{eq:barre}
  R_{21}\left(\frac{x_2}{x_1}\right)\, \bar{K_1}(x_1)\, R_{12}\left(\frac{1}{x_1 x_2}\right)\, \bar{K_2}(x_2)=
 \bar{K_2}(x_2)\,R_{21}\left(\frac{1}{x_1 x_2}\right) \,  \bar{K_1}(x_1)\,R_{12}\left(\frac{x_2}{x_1}\right)\;.
\end{equation}
The symmetry between left and right boundaries is now much more manifest
(compare \eqref{eq:re} and \eqref{eq:barre}).
\hfill\break 

 Using $R,K$ and $\widetilde K$, a one parameter  family of  transfer matrices  $t(z)$ is  defined 
  as follows \cite{sklyanin}:
\begin{equation} \label{eq:transfer_matrix}
t(z) = tr_0\Big( \widetilde K_0(z)\,T_0(z)\,K_0(z)\,\big(T_0(1/z)\big)^{-1}\Big)
\with T_0(z)=R_{0,L}(z)\,R_{0,L-1}(z)\cdots R_{01}(z).
\end{equation}
 The parameter $z$ that labels these  transfer matrices
 is known as the spectral parameter. The Yang-Baxter and the reflection
 equations imply that  $[t(x),t(y)]=0$,  for all $x,y$ \cite{Faddeev,sklyanin}.   The transfer matrices $t(z)$
 form a  one-parameter family of commuting operators that can be co-diagonalized. Their  commutation 
 ensures that  there exists enough conserved quantities to imply  integrability.

The last step  is to relate this formal construction  to
  the ASEP Markov matrix $M$. The  connection is given by 
\begin{equation}\label{eq:tpM}
M=\frac{q-1}{2} t'(1)=B_1 + \sum_{\ell=1}^{L-1} w_{\ell,\ell+1} + \bar B_L 
\end{equation}
with $B=\frac{q-1}{2} K'(1)$, $\bar B=-\frac{q-1}{2} \bar K'(1)$ and $w=(q-1)PR'(1)$,
where the prime denote derivative w.r.t. the spectral parameter $x$ and $P$ is the permutation matrix.
 The  Markov matrix $M$ belongs to the integrable  commuting family $t(z)$; it is itself integrable.
 
We observe  that the bulk transition rates are given by the choice of the $R$-matrix, 
while the  boundary matrices $B$ and $\bar B$ are obtained from the $K$-matrices. Therefore, once $R$ is given,
the problem of finding   integrable boundaries is equivalent to solving the 
reflection  and the dual-reflection equations \eqref{eq:re} and \eqref{eq:dualre}.

\subsection{Classification of the Markovian $K$-matrices for the 2-ASEP\label{sec:Kmat}}

 The $R$-matrix of the multi-species ASEP,  satisfying  the Yang-Baxter equation
 is well known \cite{Alcaraz2,Arita2}. For the 2-ASEP it is given by 
\begin{equation} \label{eq:R}
  R(x)=\left(
 \begin{array}{ccccccccc}
 1&0&0&0&0&0&0&0&0\\
 0&\frac{(x-1)q}{qx-1}&0&\frac{(q-1)x}{qx-1}&0&0&0&0&0\\
 0&0&\frac{(x-1)q}{qx-1}&0&0&0&\frac{(q-1)x}{qx-1}&0&0\\
 0&\frac{q-1}{qx-1}&0&\frac{x-1}{qx-1}&0&0&0&0&0\\
 0&0&0&0&1&0&0&0&0\\
 0&0&0&0&0&\frac{(x-1)q}{qx-1}&0&\frac{(q-1)x}{qx-1}&0\\
 0&0&\frac{q-1}{qx-1}&0&0&0&\frac{x-1}{qx-1}&0&0\\
 0&0&0&0&0&\frac{q-1}{qx-1}&0&\frac{x-1}{qx-1}&0\\
 0&0&0&0&0&0&0&0&1\\
 \end{array}
\right) .
 \end{equation}


\hfill\break 
 {\it Remark:} The ASEP $R$-matrix is a twisted version of the $U_q(\wh{gl}_3)$ $R$-matrix
  $R(x)=F_{12}\,R^{XXZ}(x)\,F_{21}^{-1}$. 
However, the twist $F_{12}$ is
\underline{not} factorisable (i.e. one cannot write $F_{12}= U_1\,V_2$). Therefore, 
 the classification \cite{K-Uq} of $K$-matrices for $U_q(\wh{gl}_3)$ cannot be used here and has to be performed 
{\it ab initio.}

\null

The problem  consists in finding the matrices  $K$ and $\widetilde K$ obeying the reflection equations
\eqref{eq:re} and  \eqref{eq:dualre}.
We focus on unitary Markovian matrices $K$ (\textit{i.e.} unitary matrices whose entries of each column sum up to $1$) such that 
the entries of the column of $B = \frac{q-1}{2} K'(1)$ sum up to $0$ (and similarly for  $\bar B$).

 The different steps to derive this classification are the following: 
 we take a generic K-matrix with 9 unknown functions as entries, we project on the 
 different entries the reflection equation \eqref{eq:re} to get 81 relations, we solve these functional equations and among the solutions we retain only  the Markovian ones.
 Altogether, we have found four different classes of solutions (each depending on $\alpha$ and $\gamma$, two 
free parameters):
\begin{equation}
K^{(1)}(x)=\left( \begin {array}{ccc}
{\frac{x((\alpha^2-\gamma^2)(x-1)+(\gamma x+\alpha)(q-1))}{(\alpha x+\gamma)((\alpha+\gamma)(x-1)+q-1)}}
&{\frac{(x^2-1)(\alpha+\gamma)\alpha}{(\alpha x+\gamma)((\alpha+\gamma)(x-1)+q-1)}}
&{\frac{(x^2-1)(\alpha+\gamma)\alpha}{(\alpha x+\gamma)((\alpha+\gamma)(x-1)+q-1)}}\\ \noalign{\medskip}
{\frac{(x^2-1)(\alpha+\gamma+1-q)\gamma}{(\alpha x+\gamma)((\alpha+\gamma)(x-1)+q-1)}}
&-{\frac{(\alpha^2-\gamma^2)(x-1)+(\alpha x+\gamma)(1-q)}{(\alpha x+\gamma)((\alpha+\gamma)(x-1)+q-1)}}
&{\frac{(x^2-1)(\alpha+\gamma)\gamma}{x(\alpha x+\gamma)((\alpha+\gamma)(x-1)+q-1)}} \\ \noalign{\medskip}
0&0&-{\frac{(\alpha+\gamma)(x-1)+x(1-q)}{x((\alpha+\gamma)(x-1)+q-1)}}
\end {array} \right)
\end{equation}

\begin{equation}
K^{(2)}(x)=\left( \begin {array}{ccc}
-{\frac{x((\alpha+\gamma)(x-1)+1-q)}{(\alpha+\gamma)(x-1)+x(q-1)}}&0&0 \\ \noalign{\medskip}
{\frac{x(x^2-1)(\alpha+\gamma)\gamma}{(\gamma x+\alpha)((\alpha+\gamma)(x-1)+x(q-1))}}
&-{\frac{x((\alpha^2-\gamma^2)(x-1)+(\gamma x+\alpha)(1-q))}{(\gamma x+\alpha)((\alpha+\gamma)(x-1)+x(q-1))}}
&{\frac{(x^2-1)(\alpha+\gamma+q-1)\gamma}{(\gamma x+\alpha)((\alpha+\gamma)(x-1)+x(q-1))}}\\ \noalign{\medskip}
{\frac{(x^2-1)(\alpha+\gamma)\alpha}{(\gamma x+\alpha)((\alpha+\gamma)(x-1)+x(q-1))}}
&{\frac{(x^2-1)(\alpha+\gamma)\alpha}{(\gamma x+\alpha)((\alpha+\gamma)(x-1)+x(q-1))}}
&{\frac{(\alpha^2-\gamma^2)(x-1)+(\alpha x+\gamma)(q-1)}{(\gamma x+\alpha)((\alpha+\gamma)(x-1)+x(q-1))}}
\end {array} \right)
\end{equation}

\begin{equation}
K^{(3)}(x)=\left( \begin {array}{ccc}
-{\frac{x((\alpha-\gamma)(x-1)+1-q)}{(\gamma x+\alpha)(x-1)+x(q-1)}}
&{\frac{(x^2-1)\gamma}{(\gamma x+\alpha)(x-1)+x(q-1)}}
&{\frac{(x^2-1)\gamma}{(\gamma x+\alpha)(x-1)+x(q-1)}}\\ \noalign{\medskip}
0&-{\frac{(\alpha x+\gamma)(x-1)+x(1-q)}{(\gamma x+\alpha)(x-1)+x(q-1)}}&0\\ \noalign{\medskip}
{\frac{(x^2-1)\alpha}{(\gamma x+\alpha)(x-1)+x(q-1)}}
&{\frac{(x^2-1)\alpha}{(\gamma x+\alpha)(x-1)+x(q-1)}}
&{\frac{(\alpha-\gamma)(x-1)+x(q-1)}{(\gamma x+\alpha)(x-1)+x(q-1)}}
\end {array} \right)
\end{equation}
and
\begin{equation}
K^{(4)}(x)=\left( \begin {array}{ccc}
{\frac{x((\gamma-\alpha)(x-1)+q-1)}{(\gamma x+\alpha)(x-1)+x(q-1)}}
&0
&{\frac{(x^2-1)\gamma}{(\gamma x+\alpha)(x-1)+x(q-1)}}\\ \noalign{\medskip}
0&1&0\\ \noalign{\medskip}
{\frac{(x^2-1)\alpha}{(\gamma x+\alpha)(x-1)+x(q-1)}}
&0
&-{\frac{(\gamma-\alpha)(x-1)+x(1-q)}{(\gamma x+\alpha)(x-1)+x(q-1)}}
\end {array} \right)
\end{equation}
They correspond respectively to the following left boundary matrices, obtained by taking the derivative at $x=1$ as in \eqref{eq:tpM}:
\begin{eqnarray}\label{solu-B}
B^{(1)} =\left( \begin {array}{ccc} -{\frac {\gamma \left(\alpha+\gamma+1-q \right) }{\alpha+\gamma}}&
\alpha&\alpha\\ \noalign{\medskip}{\frac {\gamma \left(\alpha+\gamma+1-q \right) }{\alpha+\gamma}}&-\alpha&\gamma
\\ \noalign{\medskip}0&0&-\alpha-\gamma\end {array} \right)&,& \, \, 
B^{(2)} =\left( \begin {array}{ccc} -\alpha-\gamma&0&0\\ \noalign{\medskip}\gamma&-\alpha&{\frac {
 \left(\alpha+\gamma+q-1 \right)\gamma}{\alpha+\gamma}}\\ \noalign{\medskip}\alpha&\alpha&-{\frac {
 \left(\alpha+\gamma+q-1 \right)\gamma}{\alpha+\gamma}} \end{array}\right) \\
B^{(3)} = \left( \begin {array}{ccc} -\alpha&\gamma&\gamma\\ \noalign{\medskip}0&-\alpha-\gamma&0
\\ \noalign{\medskip}\alpha&\alpha&-\gamma\end {array} \right)&,&  \,\, B^{(4)} =
\left( \begin {array}{ccc} -\alpha& 0 &\gamma\\ \noalign{\medskip}0&0&0
\\ \noalign{\medskip}\alpha&0&-\gamma\end {array} \right) \, .
\end{eqnarray}
This corresponds  using \eqref{eq:tpM} to the transition rates at the left boundary  given by
 $L_1$ to $L_4$ in the Table  \eqref{rate_left}.

The $\bar K$ matrices,  that obey \eqref{eq:barre}  can be obtained directly
 using  the formula 
\be
\bar K^{(i)}(x)=U\,K^{(i)}(1/x)\,U^{-1}\with
 U=\left(\begin{array}{ccc} 0 & 0 & 1 \\ 0 & 1 & 0 \\ 1 & 0 &0\end{array}\right),
\ee
and replacing the parameters $\alpha$ and $\gamma$ by $\beta$ and $\delta$.
The associated right boundary matrices 
$\bar B$ are given by:
\begin{eqnarray}\label{solu-Bbar}
\bar B^{(1)}=\left( \begin {array}{ccc} -\beta-\delta&0&0\\ \noalign{\medskip}\delta&-\beta&{\frac {
 \left(\beta+\delta+1-q\right)\delta}{\beta+\delta}}\\ \noalign{\medskip}\beta&\beta&-{\frac {
\left(\beta+\delta+1-q\right)\delta}{\beta+\delta}}\end {array} \right) &,& \,\, 
\bar B^{(2)}=  \left( \begin {array}{ccc} -{\frac {\delta \left(\beta+\delta+q-1\right) }{\beta+\delta}}&
\beta&\beta\\ \noalign{\medskip}{\frac {\delta \left(\beta+\delta+q-1\right) }{\beta+\delta}}&-\beta&\delta \\ \noalign{\medskip}0&0&-\beta-\delta\end {array} \right)
 \ {,} \qquad\\
\bar B^{(3)}=\left( \begin {array}{ccc} -\delta&\beta&\beta\\ \noalign{\medskip}0&-\beta-\delta&0
\\ \noalign{\medskip}\delta&\delta&-\beta\end {array} \right) &,& \,\, 
\bar B^{(4)}=\left( \begin {array}{ccc} -\delta& 0 &\beta\\ \noalign{\medskip}0&0&0
\\ \noalign{\medskip}\delta&0&-\beta\end {array} \right) 
\end{eqnarray}
where $\beta$ and $\delta$ are free parameters.
 The  transition rates at the right  boundary 
  given in Table  \eqref{rate_right} are retrieved. 
Note the exchange $q \leftrightarrow 1$ between the solutions for $B$ and those for $\bar B$.

\subsection{The $K$-matrices for the 2-TASEP}

One can perform the same classification of unitary Markovian integrable boundary conditions for TASEP models. 
The $R$-matrix for these models is the limit $q\to0$ of the ASEP $R$-matrix \eqref{eq:R}:
 \begin{equation}
  R(x)=\left(
 \begin{array}{ccccccccc}
 1&0&0&0&0&0&0&0&0\\
 0&0&0&x&0&0&0&0&0\\
 0&0&0&0&0&0&x&0&0\\
  0&1&0&1-x&0&0&0&0&0\\
  0&0&0&0&1&0&0&0&0\\
  0&0&0&0&0&0&0&x&0\\
  0&0&1&0&0&0&1-x&0&0\\
  0&0&0&0&0&1&0&1-x&0\\
  0&0&0&0&0&0&0&0&1\\
  \end{array}
 \right) 
  \end{equation}
However, the solutions to the corresponding reflection equation include unphysical solutions, where the corresponding boundary matrix $B$
has unwanted transition rates as discussed previously. 
 This unphysical solutions are exactly the cases
for which  it is impossible to construct the transfer matrix (the RHS of \eqref{eq:K_tilde} diverges when $q\to0$). Hence, in the following, we give only the solutions leading to physical 
boundary matrix $B$, which  correspond to triangular $K$ matrices.
The triangular Markovian $K$-matrices solutions to the reflection equation are given by
\begin{equation}
K^{(1)}(x)=\left( \begin {array}{ccc}
\frac{x(\mu (x-1)+1)}{(1-\mu)x+\mu}&0&0 \\ \noalign{\medskip}
-\frac{\mu(x^2-1)}{(1-\mu)x+\mu}&1&0\\ \noalign{\medskip}
0&0&1
\end {array} \right)\,,\qquad
K^{(2)}(x)=\left( \begin {array}{ccc}
x^2&0&0 \\ \noalign{\medskip}
-{\frac{x(x^2-1)(\alpha-1)}{\alpha(x-1)-x}}&-{\frac{x(\alpha(x-1)+1)}{\alpha(x-1)-x}}&0\\ \noalign{\medskip}
{\frac{\alpha(x^2-1)}{\alpha(x-1)-x}}&{\frac{\alpha(x^2-1)}{\alpha(x-1)-x}}&1
\end {array} \right)
\label{eq:KTASEP}
\end{equation}
\begin{equation}
K^{(3)}(x)=\left( \begin {array}{ccc}
-{\frac{x(\alpha(x-1)+1)}{\alpha(x-1)-x}}&0&0\\ \noalign{\medskip}
0&-{\frac{x(\alpha(x-1)+1)}{\alpha(x-1)-x}}&0\\ \noalign{\medskip}
{\frac{\alpha(x^2-1)}{\alpha(x-1)-x}}&{\frac{\alpha(x^2-1)}{\alpha(x-1)-x}}&1
\end {array} \right)
\,,\qquad
K^{(4)}(x)=\left( \begin {array}{ccc}
-{\frac{x(\alpha(x-1)+1)}{\alpha(x-1)-x}}&0&0\\ \noalign{\medskip}
0&1&0\\ \noalign{\medskip}
{\frac{\alpha(x^2-1)}{\alpha(x-1)-x}}&0&1
\end {array} \right)
\end{equation}

They correspond respectively to the following left boundary matrices:
\begin{equation}\label{solu-B_TASEP}
B =\left( \begin {array}{ccc}
-\mu&0&0\\ \noalign{\medskip}
\mu&0&0\\ \noalign{\medskip}
0&0&0
\end {array} \right)
\mb{,}\left( \begin {array}{ccc}
-1&0&0\\ \noalign{\medskip}
1-\alpha&-\alpha&0\\ \noalign{\medskip}
\alpha&\alpha&0
\end {array} \right)
\mb{,}
\left( \begin {array}{ccc} 
-\alpha&0&0\\ \noalign{\medskip}
0&-\alpha&0\\ \noalign{\medskip}
\alpha&\alpha&0
\end {array} \right)
 \mb{,}
\left( \begin {array}{ccc} 
-\alpha&0&0\\ \noalign{\medskip}
0&0&0\\ \noalign{\medskip}
\alpha&0&0
\end {array} \right) 
\end{equation}
where $\alpha$, $\mu$ are free parameters. The transition rates
 given in  Table \eqref{rate_left_t} are thus  obtained.
Using the same symmetry as for the ASEP case, we can construct the $\bar K$ matrices
 and  the  $\bar B$ operators: 
\begin{equation}\label{solu-Bbar_TASEP}
\bar B =\left( \begin {array}{ccc}
0&0&0\\ \noalign{\medskip}
0&0&\nu\\ \noalign{\medskip}
0&0&-\nu
\end {array} \right)
\mb{,}
\left( \begin {array}{ccc}
0&\beta&\beta\\ \noalign{\medskip}
0&-\beta&1-\beta\\ \noalign{\medskip}
0&0&-1
\end {array} \right)
\mb{,}
\left( \begin {array}{ccc} 
0&\beta&\beta\\ \noalign{\medskip}
0&-\beta&0\\ \noalign{\medskip}
0&0&-\beta
\end {array} \right)
 \mb{,}
\left( \begin {array}{ccc} 
0&0&\beta\\ \noalign{\medskip}
0&0&0\\ \noalign{\medskip}
0&0&-\beta
\end {array} \right) 
\end{equation}
where $\beta$, $\nu$ are free parameters. This corresponds to
 the transition rates given in Table  \eqref{rate_right_t}.

\subsection{Matrix Ansatz\label{sec:MA}}

 We now apply the results of \cite{CRV} to construct a matrix product
 representation of the steady state using the  fundamental  $R$,  $K$
 and $\bar{K}$ matrices found above. We 
 believe that integrability provides a natural framework to obtain a 
 bona fide  matrix ansatz.   As explained in \cite{CRV},  this claim  is based
on the analysis of  ZF algebras  and the study of their
   representations as a  starting point.
In  section \ref{example}, we shall give an example 
 of a  2-TASEP for which integrability
 leads to  a matrix ansatz.

More precisely, we  define a  $3$-component column vector $A(x)$
depending on a spectral parameter $x$ and with algebraic entries,
satisfying the following equations: 

 \bea
\label{eq:ZF}
&& R_{12}\left(\frac{x_1}{x_2}\right)A_1(x_1)A_2(x_2)=A_2(x_2)A_1(x_1).
\\
 \label{eq:GZ}
&& \llangle W|\left(K(x)A\left(\frac{1}{x}\right)-A(x)\right)=0, \quad \text{ and } \quad 
 \left(\bar K(x)A\left(\frac{1}{x}\right)-A(x)\right)|V\rrangle =0.
\eea
 Equation \eqref{eq:ZF}, valid in the bulk, is  known as the
 Zamolodchikov-Faddeev (ZF) relation \cite{ZF}.  The boundary equations of
 \eqref{eq:GZ} are known as  Ghoshal-Zamolodchikov (GZ) relations \cite{GZ}.
 These equations  allow us to construct  a matrix Ansatz in the sense
 of \cite{DEHP}.  The  link between ZF  and the matrix product
 representation  was first found   by \cite{Sasamoto2} and   the
 precise  relation between ZF  and GZ algebras and  the  matrix Ansatz
 was given  in   \cite{CRV}. The key point to derive the matrix ansatz  from the ZF algebra and the GZ relations is to take the derivative of \eqref{eq:ZF} at $x_1=x_2=1$ and of \eqref{eq:GZ} at $x=1$. This  leads to
 \bea
&& w\,A_1(1)A_2(1)=(q-1)\Big(A_1(1)A_2'(1)-A_1'(1)A_2(1)\Big),
\label{ZF-derivee}\\
&& \llangle W|\Big(BA(1)-(q-1)A'(1)\Big)=0, \quad \text{ and } \quad 
 \Big(\bar BA(1)+(q-1)A'(1)\Big)|V\rrangle =0. \label{GZ-derivee}\
 \eea
 Using these relations, on can show \cite{CRV} that the vector defined by
\begin{equation}
 \steady = \frac{1}{Z_L}\llangle W| A_1(1)\dots A_L(1)|V\rrangle,
\with Z_L=\llangle W|C^L|V\rrangle \and C=(1,1,1).A(1),
\label{ZFGZDEHP}
\end{equation}
is the steady state of the  process, \textit{i.e.} $M\steady=0$.
Writing the components of this vector $\steady$,  the weights of any configuration at the stationary state
 are obtained.  

 The crucial step is to construct  the  vector $A(x)$. As explained in \cite{CRV}, this vector 
 is a Laurent series with respect to
 the  spectral parameter $x$, but this series can be truncated and only a finite number of terms retained.
 We shall illustrate this procedure in the next section, devoted to the analysis of a specific example.

\section{A specific example \label{example}}

\begin{figure}[htb]
\begin{center}
 \begin{tikzpicture}[scale=0.7]
\draw (-4,0) -- (14,0) ;
\foreach \i in {-4,-3,...,14}
{\draw (\i,0) -- (\i,0.4) ;}
\draw[->,thick] (-4.4,0.9) arc (180:0:0.4) ; \node at (-4.,1.8) [] {$\frac{1}{2}$};
\draw[<-,dashed,thick] (-3.6,-0.1) arc (0:-180:0.4) ; \node at (-4.,-1.) [] {$\frac{1}{2}$};
\draw  (-1.5,0.5) circle (0.3) [circle] {};
\draw  (1.5,0.5) circle (0.3) [fill,circle] {};
\draw  (4.5,0.5) circle (0.3) [fill,circle] {};
\draw  (5.5,0.5) circle (0.3) [fill,circle] {};
\draw  (6.5,0.5) circle (0.3) [circle] {};
\draw  (9.5,0.5) circle (0.3) [circle] {};
\draw  (10.5,0.5) circle (0.3) [fill,circle] {};
\draw  (13.5,0.5) circle (0.3) [fill,circle] {};
\draw[->,thick] (-1.6,1) arc (0:180:0.4); \draw[thick] (-1.8,1.2) -- (-2.2,1.6) ; \draw[thick] (-1.8,1.6) -- (-2.2,1.2) ;
\draw[->,thick] (-1.4,1) arc (180:0:0.4); \node at (-1.,1.8) [] {$1$};
\draw[->,thick] (1.4,1) arc (0:180:0.4); \draw[thick] (1.2,1.2) -- (0.8,1.6) ; \draw[thick] (1.2,1.6) -- (0.8,1.2) ;
\draw[->,thick] (1.6,1) arc (180:0:0.4); \node at (2.,1.8) [] {$1$};
\draw[<->,thick] (5.6,1) arc (180:0:0.4); \node at (6.,1.8) [] {$1$};
\draw[->,thick] (6.6,1) arc (180:0:0.4); \node at (7.,1.8) [] {$1$};
\draw[<->,thick] (9.6,1) arc (180:0:0.4); \draw[thick] (9.8,1.2) -- (10.2,1.6) ; \draw[thick] (9.8,1.6) -- (10.2,1.2) ;
\draw[->,thick] (10.6,1) arc (180:0:0.4); \node at (11.,1.8) [] {$1$};
\draw[->,thick] (13.6,1) arc (180:0:0.4) ; \node at (14.,1.8) [] {$1$};
 \end{tikzpicture}
 \end{center}
 \caption{The 2-TASEP with special, integrable, boundary rates. The empty sites stands for species $1$,
  circles for species $2$ and  bullets
 for species $3$. On the left boundary the continuous line means injection of bullets whereas the dashed line means injection of circles. \label{fig:rdm}}
\end{figure}
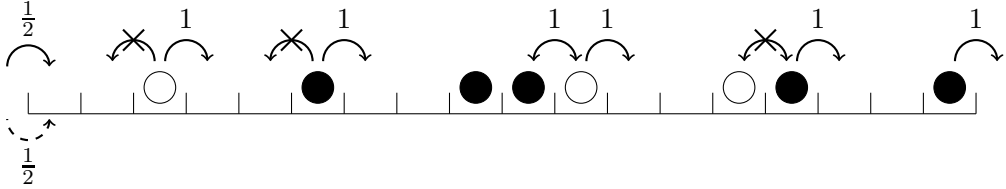

   In this section, we shall consider a specific example for which
 we shall carry out a more detailed analysis. 
We consider a 2-TASEP and   take the set of transition rates $L_2$ with $\alpha=1/2$  
and the set of transition rates $R_3$ with $\beta=1$ 
 (see Figure \ref{fig:rdm}).
 In brief, the example we consider is given by the following rules: 
 \begin{equation}
 \begin{array}{|c |c| c| }
 \hline \text{Left} & \text{Bulk} & \text{Right} \\
 \hline
 1\, \xrightarrow{\ 1/2\ }\, 2&  21\, \xrightarrow{\ 1\ } \,12&2\, \xrightarrow{\ 1\ } \,1\\
 1\, \xrightarrow{\ 1/2\ }\, 3&31\, \xrightarrow{\ 1\ }\, 13&3\, \xrightarrow{\ 1\ }\, 1\\
 2\, \xrightarrow{\ 1/2\ } \,3&32\, \xrightarrow{\ 1\ } \,23&\\ \hline
 \end{array}
 \end{equation}
  Note  that for these rules,
 the boundaries are permeable for all  the species.

 We  now  construct an algebra  which allows us to 
compute the weights of the stationary state. We will then  give several quantities computed with this Ansatz.

The stationary state of the process can be expressed in a matrix product form, {\it i.e.},
the weight of the configuration $\mathcal{C}=(\tau_1,\dots,\tau_L)$ in the stationary state can be written as 
\begin{equation} \label{eq:weight}
 P(\tau_1,\dots,\tau_L)=\frac{1}{Z_L}\llangle W| X_{\tau_1}X_{\tau_2}\cdots X_{\tau_L} |V\rrangle,
\end{equation}
where $Z_L=\llangle W|(X_1+X_2+X_3)^L|V\rrangle$ is the normalization factor.
 Using   \eqref{ZFGZDEHP}, we find that the  operators $X_1,X_2$ and $X_3$
 are given  by
\begin{equation} \label{XfromA}
  A(1)=\left(\begin{array}{c}
 X_1\\
 X_2\\
 X_3
 \end{array}\right)\;, 
\end{equation}
 where the vector $A(x)$  satisfies the ZF and GZ relations \eqref{eq:ZF} and \eqref{eq:GZ}. 
The exchange relations among the $X_i$'s are obtained from relations \eqref{ZF-derivee}. However, the generators that appear in  $A'(1)$ are not necessarily scalars 
 (they will {\it not} be  scalars in the present case,  see below). 
 Therefore,  more relations are required to close the algebra generated by the $X_i$'s. 
 A systematic way to deal with this question \cite{CRV}  is to assume 
 the following  expansion for the vector $A(x)$:
\begin{equation} \label{dev_A}
 A(x)=\left(\begin{array}{c}
 x^2+G_9x+G_8+G_7/x\\
 G_6x+G_5+G_4/x\\
 G_3x+G_2+G_1/x+1/x^2
 \end{array}\right)\;, 
\end{equation}
where the $G_i$'s belong to a non-commuting algebra. 
We have observed that  9 generators is the minimal choice  that ensures
 that no word of length 3  built from the $X_i$'s
identically vanishes. A general proof of this fact
 for words of arbitrary length is still missing. 
 A  way to proceed would be to construct an explicit representation of the  algebra
 generated by the $X_i$'s.  
We have obtained some partial results, representing the $X_i$'s in terms of tensor products of DEHP generators \cite{CMRV}, but the general scheme remains to be found.
 Below, we give  general formulas for some
 special  words of arbitrary length.

From  \eqref{XfromA}, we have 
\begin{equation}\begin{aligned}
 X_1 &= 1+G_9+G_8+G_7 \\
 X_2 &= G_6+G_5+G_4 \\
 X_3 &= 1+G_3+G_2+G_1
\end{aligned}\label{XenG}
\ee
 The principle is the similar  as the one introduced in \cite{DEHP}
 except that in our case the algebra is more involved since it  
 contains $9$  fundamental generators $G_1,G_2,\dots,G_9$ from which the three 
 matrices, corresponding to the three types of particles,   $X_1,X_2$ and $X_3$ are expressed. 

 The algebra satisfied by  the nine generators $G_i$ is found  by writing each 
component of the ZF relation and identifying the  coefficients of the polynomials in $x_1$ and $x_2$.
The  generators $G_i$ satisfy a quadratic algebra, given by the following {\it exchange} relations:
 \begin{align}
&\left[G_1,G_2\right]=0, \nonumber\\
&\left[G_1,G_3\right]=0, &&\left[G_2,G_3\right]=0,\nonumber\\
&\ G_1G_4=G_5, &&\ G_2G_4=G_6, &&\ G_3G_4=0,\nonumber \\
&\left[G_1,G_5\right]= G_6-G_4G_2,&&\  G_2G_5=G_1G_6, &&\ G_3G_5=0, && \left[G_4,G_5\right]=0,\nonumber\\
&\left[G_1,G_6\right]= -G_4\,G_3, && \left[G_2,G_6\right]= -G_5G_3, &&\ G_3G_6=0, && \left[G_4,G_6\right]=0,\nonumber\\
&\  G_1G_7=G_8, &&\ G_2G_7=G_9, &&\ G_3G_7=1, &&\ G_4G_7=0,\nonumber \\
&\left[G_1,G_8\right]=G_9-G_7G_2, &&\ G_2G_8=G_1G_9, &&\ G_3G_8=G_1, &&\left[G_4,G_8\right]=-G_7G_5,\nonumber\\
&\left[G_1,G_9\right]=1-G_7G_3, &&\left[G_2,G_9\right]=G_1-G_8G_3, &&\ G_3G_9=G_2, &&\left[G_4,G_9\right]=-G_7G_6,\nonumber\\
&\left[G_5,G_6\right]=0,\label{rel_exchange} \\
&\ G_5G_7=0, &&G_6G_7=0, \nonumber\\
& \ G_5G_8=G_4G_9,&&G_6G_8=G_4,&&\left[G_7,G_8\right]=0,\nonumber  \\
& \left[G_5,G_9\right]=G_4-G_8G_6,&&G_6G_9=G_5,&&\left[G_7,G_9\right]=0,&&\left[G_8,G_9\right]=0.\nonumber
 \end{align}
\hfill\break
Remark: From the knowledge of the  exchange relations for the $G_i$'s, the
 algebra generated by the $X_i$'s can  be obtained  using \eqref{XenG}
 and  \eqref{rel_exchange}. 
We need six more generators  $\bar X_i$ and 
$\bar{\bar{X_i}}$ corresponding to 
\be
A'(1)=\left(\begin{array}{c}
 \bar X_1\\
\bar X_2\\
\bar X_3
 \end{array}\right)\,,\quad A''(1)=\left(\begin{array}{c}
\bar{\bar{X_1}}\\
\bar{\bar{X_2}}\\
\bar{\bar{X_3}}
 \end{array}\right).
 \ee
 Note that these generators are not scalar. 
 The  algebra generated by $X_i$, $\bar X_i$ and $\bar{\bar{X_i}}$ is  the same as the one generated by the $G_i$'s, written in a different  basis. We  chose to present 
 the commutation relations in the $G_i$ basis because they are simpler.

The action of the  $G_i$'s on the boundary vectors
 is derived  from the GZ relations \eqref{eq:GZ}:
\begin{equation} \label{rel_boundaries}
  \begin{array}{ll}
 \llangle W|\left(G_4-1\right)=0, \hspace{1cm}& G_3\,|V\rrangle =0,\\
 \llangle W|\,G_7=0, \hspace{1cm}& G_5\,|V\rrangle =0,\\
 \llangle W|\left(G_8-1\right)=0,\hspace{1cm}& G_6\,|V\rrangle =0, \\
 \llangle W|\left(G_1-G_3-G_5\right)=0, \hspace{1cm}& \left(G_2-1\right)|V\rrangle =0,\\
 \llangle W|\left(G_6+G_9-G_5-1\right)=0, \hspace{1cm}& \left(G_1+G_4+G_7-G_9\right)|V\rrangle =0.
  \end{array}
 \end{equation}
 Note that the relations are not all independent. 
For instance, $G_3G_5=0$ and $\left(G_2-1\right)|V\rrangle =0$ can be deduced from 
the other exchange and boundary relations.

  Using the  algebraic relations \eqref{rel_exchange}-\eqref{rel_boundaries},  any expression containing 
  $X_1$, $X_2$ and $X_3$ and placed  between $\llangle W|$ and $|V\rrangle$
  can be reduced to a multiple  of $\llangle W|V\rrangle$.
Therefore,  the weights of any configuration  and the partition function of the model
  can be calculated. A computer program allowed us to obtain
 all  the  weights for systems of small sizes.  In the rest of this section, we
 present some  results and defer technical details and proofs  to 
 a forthcoming  publication \cite{CMRV}. 
 
 $\bullet$ A  word containing only  the letters $X_2$ and $X_3$ is evaluated as 
\begin{equation}
 \llangle W|Y^{(k)}(X_2,X_3)X_3^p|V\rrangle = \frac{k+2}{p+k+2}
 \left(
 \begin{array}{c}
 2p+k+1\\
 p
 \end{array}
\right) \llangle W|V\rrangle,\qquad \text{for } p,\,k=0,1,2,\dots,
\end{equation}
where $Y^{(k)}(X_2,X_3)$ is the empty word for $k=0$ and is a word of length $k$ containing only 
 the letters $X_2$ and $X_3$ and ending with $X_2$.

$\bullet$ A word with the letter $X_1$ on the left is calculated as
\begin{equation}
 \llangle W|X_1^p\,Y^{(k)}(X_2,X_3)|V\rrangle = \frac{2p+1}{p+1} \left( \begin{array}{c}
 2p \\p \end{array} \right)\,
 \llangle W|V\rrangle,\qquad \text{for } p,\,k=0,1,2,\dots,
\end{equation}
where $Y^{(k)}(X_2,X_3)$ is as above.

 $\bullet$    The partition function is given by 
\be
 Z_L=\llangle W|C^L |V\rrangle=(2L+1)A_L A_{L+1} \llangle W|V\rrangle,
\ee
where $C=X_1+X_2+X_3$ and $A_L=\frac{1}{L+1}\left( \begin{array}{c}
 2L \\L
 \end{array} \right)$ is the Catalan number.

 $\bullet$ The average density of  particles of type $i$  at site $k$ is given by
 $ n_i^{(k)}=\frac{1}{Z_L}\llangle W|C^{k-1}X_iC^{L-k}|V\rrangle .$
We obtain:
\begin{eqnarray}
 n_1^{(k)} &=& \frac{1}{A_{L+1}}\sum_{i=0}^{k-1}A_iA_{L-i}, 
\\ n_2^{(k)} &=& \frac{1}{A_{L+1}}\sum_{i=k}^{L}\frac{L-i+1}{L+2}A_iA_{L-i},  \\
 n_3^{(k)} &=& \frac{1}{A_{L+1}}\sum_{i=k}^{L}\frac{i+1}{L+2}A_iA_{L-i}\;.
\end{eqnarray}
 
$\bullet$  The current $j_i$ of the particles of type $i$ 
is defined  as
\begin{eqnarray}
 &&j_1=-\frac{1}{Z_L}\llangle W|C^{k-1}(X_2+X_3)X_1C^{L-k-1}|V\rrangle, \\
 &&j_2=\frac{1}{Z_L}\llangle W|C^{k-1}(X_2X_1-X_3X_2)C^{L-k-1}|V\rrangle, \\
 &&j_3=\frac{1}{Z_L}\llangle W|C^{k-1}X_3(X_1+X_2)C^{L-k-1}|V\rrangle.
\end{eqnarray}
Using algebraic relations \eqref{rel_exchange}-\eqref{rel_boundaries} we can show that 
the currents are independent of $k$.
They have  the following simple expressions
\begin{equation}
 j_1 = -\frac{L+2}{2(2L+1)}, \quad j_2 = \frac{1}{2(2L+1)} \ \ \text{ and } \ \ j_3 = \frac{L+1}{2(2L+1)}.
\end{equation}

 We note that  densities and  particle currents can be computed using
 the identification  method of   \cite{Ayyer}, a procedure  that can
 also be well-defined at the level of the algebra \cite{CMRV}.
 However, the algebraic relations
 \eqref{rel_exchange}-\eqref{rel_boundaries} allow us to compute all
 the individual weights and correlations between different type of
 particles: this cannot be obtained from  the identification
 procedure.  Finally, we mention  that similar  algebras, with at most
 nine generators,  can be defined to compute the weight of the
 stationary state of the other integrable 2-ASEP  model found in this
 work. 

\section{Conclusion}

  In this work,  we give a complete classification of the integrable unitary Markovian
  boundary conditions for the 2-ASEP with open boundaries. This
  classification is obtained by solving the Sklyanin  reflection
  equations for the boundary $K$-matrices, and then demanding the solutions to be unitary and Markovian. We emphasize that for some
  choices of these boundary conditions, all  particle currents are
  non-vanishing. Hence the system is genuinely  out-of-equilibrium and
  can  exhibit phase transitions when the boundary parameters are varied.
  We claim that the stationary state of all these  models can  be
  built in a matrix product form. This statement is illustrated by a
  detailed study of a particular  open 2-TASEP model. For this model,
  we construct  explicitly a matrix Ansatz algebra by  using two  key
  ingredients,  the Zamolodchikov-Faddeev  (ZF) and  the
  Ghoshal-Zamolodchikov (GZ) relations. The algebra has nine
  generators and two boundary vectors;  the  ZF relation gives the
  bulk exchange relations between these  generators  whereas  GZ
  gives  their  action  on the  two boundary vectors.  It can be shown
  that these rules allow us to calculate all stationary weights.
  Several quantities are computed thanks to  this algebra,  revealing
  the beautiful combinatorial structure that underlies this integrable
  model.  

   Many problems remain to be solved.  We have not performed here  an
   exhaustive analysis of all the types of boundaries obtained in the
   classification. Some of the models may display a rich
   phenomenological behaviour and  deserve further study \cite{CMRV}.
   Besides, it would be useful to  relate the algebra generated  from
   ZF and GZ with the tensor products of $q$-deformed oscillator
   algebras used in  \cite{PEM} to study the $N$-ASEP (it is plausible
   that these  tensor products give  explicit representations of the
   former algebra).  More generally, integrability  should allow us
   to calculate various  physical quantities;   for example, the
   spectral gap of  the standard ASEP is known \cite{DE}. Can  one
   determine  the gap of the 2-ASEP?    Recent  progress has  been
   made in computing the eigenvectors and the eigenvalues using the
   algebraic Bethe Ansatz for integrable models with generic
   boundaries \cite{BCR,BC,Bel,crampe13}. These techniques  should  be
   generalized to the models investigated  here.

  Finally,  a systematic  investigation of  integrable boundaries for
  various generalizations of  the  exclusion processes remains an open
  problem. Some  particularly  interesting cases are the following:
  the 2-TASEP model with unequal  hopping rates \cite{ALS2} (the bulk
  dynamics is known to be  integrable \cite{Cantini}),  the homogeneous
  N-species ASEP (for   generic N  a brute force resolution of the
  reflection equation seems  difficult but partial results
  were  obtained using a baxterisation of a boundary Hecke algebra
  \cite{Cantini-P}) and  the bridge model  which is variant of
  the 2-TASEP.  The bridge  model, which for some choices of boundary rates displays a 
   fascinating spontaneous  symmetry-breaking \cite{Evans,Gianni}, 
  has eluded an exact solution. We hope that the formalism
  developed here will lead to some progress in solving  this problem.

\end{document}